\documentclass[%
 reprint,
 amsmath,amssymb,
 aps,prc
]{revtex4-2}
\usepackage[colorlinks, citecolor=red]{hyperref}
\usepackage{graphicx}
\usepackage{dcolumn}
\usepackage{bm}
\usepackage{nameref}
\usepackage{natbib}
\usepackage[T1]{fontenc}
\usepackage{booktabs, array, mathptmx, float, tabularx, booktabs, lipsum, amsmath,multirow}
\usepackage{siunitx, xcolor}
\usepackage[version=4]{mhchem}
\begin{document}
\preprint{APS/123-QED}
\title{Nonlocality effect in $\alpha$ decay half-lives for even-even nuclei within a two potential approach}

\author{ Jinyu Hu$^{1}$ and  Chen Wu$^{1}$ } \affiliation{
\small 1. Xingzhi College, Zhejiang Normal University, Jinhua, 321004, Zhejiang, China}
\begin{abstract}
  In this paper, we carefully look at the \texorpdfstring{$\alpha$}{}-decay half-lives of 196 even-even nuclei using a two-potential approach that is made better by taking into account an alpha particle's effective mass that changes with coordinates. The result shows that the accuracy of this model has been improved after considering effective mass for the alpha particle.
Furthermore, considering $\alpha$ decay energies derived from three mass models, namely the Weizsacker-Skyrme-4 (WS4) mass model, the relativistic continuum Hartree-Bogoliubov theory mass model, and the FRDM (2012), we extend this model to predict the \texorpdfstring{$\alpha$}{}-decay half-lives of Z = 118 and 120 isotopes. Finally, we carefully study the predicted $\alpha$ decay energies and half-lives of Z = 118 and 120 isotopes and discuss the shell structure of superheavy nuclei. We found that the shell effect is obvious at N = 178 and at N = 184 in the WS4 mass model and FRDM(2012), while the shell effect is only obvious at N = 184 in the relativistic continuum Hartree-Bogoliubov theory mass model.
\end{abstract}

\maketitle

\section{\label{sec:level1}INTRODUCTION}
In 1899, Rutherford made the pioneering discovery of the natural emission of helium nuclei. Subsequently, in 1911, Gurney, Condon \cite{gurney1928wave} and Gamow \cite{gamow1928quantentheorie} independently provided theoretical explanations for this phenomenon grounded in the quantum tunneling theory. Since then, $\alpha$ decay has emerged as a focal point of research in nuclear physics. This is primarily because $\alpha$ decay serves as a valuable probe, offering profound insights into various aspects of nuclear structure. For instance, it provides crucial information regarding shell closure effects, nuclear level structures, and ground-state properties of atomic nuclei, among other fundamental characteristics \cite{matsuse1975study, wauters1994fine, kucuk2020role, wang2017competition, xiao2020alpha}.
\par{}
Despite several decades of development in this field, the accurate theoretical interpretation of \texorpdfstring{$\alpha$}{}-decay remains an outstanding challenge. Furthermore, with the continuous advancement of technology and experimental measurement techniques, an increasing number of exotic structures of unstable nuclei and chemical elements with high atomic numbers \cite{wang2021ame, kondev2021nubase2020, terranova2022periodic} have been identified. For instance, in 2011, the superheavy element Z = 118 was successfully synthesized by conducting $^{48}$Ca-induced hot fusion reactions using the actinide element Cf as the target material \cite{oganessian2011synthesis}.
\par{}
In an effort to reduce the disparity between the computed $\alpha$ decay half-lives and the corresponding experimental data, a plethora of investigations into empirical formulas and theoretical frameworks for $\alpha$ decay have been carried out. Such as Coulomb and proximity potential model \cite{santhosh2012cluster}, density-dependent cluster model \cite{xu2005favored, xu2005systematical}, universal decay law \cite{qi2009universal}, Royer formula \cite{royer2000alpha}, Gamow-like model \cite{zdeb2013half}, two-potential approach (TPA) \cite{gurvitz1987decay} and others \cite{jian2009alpha, yan2009branching}. These studies are aimed at filling the gaps in the existting $\alpha$-decay theory and better interpreting the new experimental findings of superheavy and exotic-structured nuclei.
\par{}
Superheavy nuclei represent another significant research focus within the realm of alpha decay, garnering substantial attention from the scientific community. A multitude of theoretical model - based studies have been conducted in this area.\cite{sobiczewski1989deformed,rutz1997superheavy, maglione1998nucleon, denisov2009alpha, hamilton2013search, zagrebaev2001synthesis, smolanczuk1997properties, sobiczewski2007description, zagrebaev2008synthesis, xu2019alpha}. Recently, E.L.Medeiros et al \cite{medeiros2022nonlocality} systematically studied  $\alpha$ decay half-lives with  52 $\le $ Z $\le $ 103 considering effective mass for the alpha particle. Their result indicated that the nonlocality effect has a certain effect on $\alpha$ decay half-lives.
In this study, we aim to extend the semi-classical WKB phenomenological approach, which incorporates a coordinate-dependent effective mass for the alpha particle, to the TPA. Subsequently, we utilize this modified TPA to calculate the \texorpdfstring{$\alpha$}{}-decay half-lives of 196 even-even nuclei.For the purpose of comparison, the decay energies were calculated using three distinct mass models,namely the Weizsacker-Skyrme-4 (WS4) mass model, the relativistic continuum Hartree-Bogoliubov theory mass model, and the FRDM (2012), respectively.
\par{}
The article is organized as follows. The theoretical framework of the TPA and the effective mass for the alpha particle are briefly presented in section. The detailed calculations and discussion are given in section. A summary is given in section

\section{\label{sec:level2}THEORETICAL FRAMEWORK }
 \subsection{TPA framework}\label{A}
The half-life $T_{1/2}$ is an important indicator for nuclear stability, which can be calculated by the decay width
$\Gamma $ written as:
\begin{equation}\label{eq1}
T_{1/2}= \frac{\hbar ln 2}{\Gamma}
\end{equation}
where $\hbar$ is the reduced Planck constant. In the framework of TPA \cite{gurvitz2004modified,gurvitz1987decay}, the $\alpha$ decay width $\Gamma$ can be expressed as:
\begin{equation}\label{eq2}
\Gamma= \frac{\hbar^{2}P_{\alpha} F P}{4\mu}
\end{equation}
where $\mu$ is the reduced mass of the $alpha$-daughter nucleus system. $F$ is the normalized factor, expressing the $\alpha$ particle assault frequency, can be given by
\begin{equation}\label{eq3}
F= \frac{1}{\int_{r_{1}}^{r_{2}}\frac{1}{2k(r)}dr }
\end{equation}
Here $k(r) = \sqrt{(\frac{2\mu}{\hbar^{2}}\left | Q_{\alpha} - V(r) \right | )}$ represents the wave number of the $\alpha$ particle. The $\alpha$-core $V(r)$ is the total interaction potential between the daughter nucleus and the emitted $\alpha$ particle. $Q_{\alpha}$ is the $\alpha$ decay energy. The semiclassical Wentzel-Kramers-Brillouin (WKB)  barrier penetration probability $P$ can be written by \cite{ismail2017alpha}
\begin{equation}\label{eq4}
P=exp \left [  -2 \int_{r_{2}}^{r_{3}}k(r)dr  \right ]
\end{equation}
where $r_{1}$, $r_{2}$, and $r_{3}$ are the classical turning point and satisfy the conditions $V(r_{1}) = V(r_{2}) = V(r_{3}) = Q_{\alpha}$. In the parent nucleus ($r_{1}$ < $r$ < $r_{2}$), the corresponding interactions are dominated by the nuclear potential, whereas, in the region outside the nucleus ($r_{2}$ < $r$ < $r_{3}$), the Coulomb potential has an important role.
\par{}
The preformation probability $P_{\alpha}$ can be calculated from the the cluster formation model. In 2013,  Ahmed et al \cite{ahmed2013alpha, ahmed2013clusterization} proposed to calculate the preformation probability of even-even heavy nuclei. In this work, the preformation probability $P_{\alpha}$ can be  calculated by
\begin{equation}\label{eq5}
P_{\alpha}=\frac{E_{f\alpha}}{E}
\end{equation}
Where $E_{\alpha}$ is $\alpha$ cluster-formation energy and $E$ is total energy. For even-even nuclei they can be given by \cite{ahmed2013clusterization,deng2018systematic}
\begin{equation}\label{eq6}
\begin{aligned}
E_{f\alpha} = & 3B(A,Z)+B(A-4,Z-2)   \\&
                   -2B(A-1,Z-1)-2B(A-1,Z),
\end{aligned}
\end{equation}
\begin{equation}\label{eq7}
E=B(A,Z)-B(A-4,Z-2)
\end{equation}
Here $B(A,Z)$ is the binging energy of the nucleus with the mass number $A$ and proton number $Z$.
\par{}
The $\alpha$-core potential $V(r)$, including the nuclear potential $V_{N}(r)$, Coulomb potential $V_{C}(r)$, and centrifugal potential $V_{l}(r)$, is written as
\begin{equation}\label{eq8}
V(r)=V_{N}(r)+V_{C}(r)+V_{l}(r)
\end{equation}
\par{}
In this work, the nuclear potential $V_{N}(r)$ is described by the type of cosh parametrized form \cite{buck1992alpha}. It is written as
\begin{equation}\label{eq9}
V_{N}(r)=-V_{0}\frac{1+cosh(R/a)}{cosh(r/a)+cosh(R/a)}
\end{equation}
where $V_{0}$ and $a$ are parameters of the depth and diffuseness of the nuclear potential, respectively. In Xiao-Hua Li's study,
we have obtained  a series of parameter value, i.e., $a = 0.5958$ fm and $V_{0} = 192.42 +31.059 \frac{N-Z}{A}$ MeV \cite{sun2016systematic}, where $N$, $Z$, and $A$ represent the neutron, proton, and mass number of daughter nucleus, respectively. $V_{C}(r)$ represents the Coulomb potential, which is taken as the potential of a uniformly charged sphere with sharp radius $R$ and can be expressed as
\begin{equation}\label{eq10}
V_{C}(r)  =\left\{\begin{matrix}
 \frac{Z_{d}Z_{\alpha}e^{2}}{2R}\left [  3 - \frac{r^{2}}{R^{2}}\right ] & r\le R \\
  \frac{Z_{d}Z_{\alpha}e^{2}}{r},& r > R,
\end{matrix}\right.
\end{equation}
where $Z_{d}$ and $Z_{\alpha}$ denote the charge number of the daughter nucleus and $\alpha$-particle, respectively. The radius $R$ is written as:
\begin{equation}\label{eq11}
R=1.28 A^{1/3} - 0.76+ 0.8A^{-1/3}
\end{equation}
\par{}
Because $l(l+1)$ $\to $ $(l+\frac{1}{2})^{2}$ is necessary correction for one-dimensional problems\cite{morehead1995asymptotics}, in this work, we adopt the Langer modified centrifugal barrier $V_{l}(r)$. It can be exprssed as
\begin{equation}\label{eq12}
V_{l}(r)=\frac{\hbar^{2}(l+\frac{1}{2})^{2}}{2\mu r^{2}}.
\end{equation}
\par{}
The minimum angular momentum \cite{denisov2009alpha} $l_{min}$ obtained by the $\alpha$-particle is selected in based on the conservation laws of spin-parity, which can be written as\cite{sun2017systematic}
\begin{equation}\label{eq13}
l_{min}= \left\{\begin{matrix}
 \Delta_{j}, &for  &even  & \Delta_{j} &and  &\pi_{p}=\pi_{d}, \\
  \Delta_{j+1},& for &even  & \Delta_{j} &and  & \pi_{p}\ne \pi_{d},\\
  \Delta_{j},&  for&  odd& \Delta_{j} & and & \pi_{p}=\pi_{d},\\
  \Delta_{j+1},&  for&  odd& \Delta_{j} & and &\pi_{p}\ne \pi_{d},
\end{matrix}\right.
\end{equation}
where $\Delta_{j} = \left | j_{p} - j_{d} \right |$, $j_{p}$, $\pi_{p}$, $j_{d}$, $\pi_{d}$ represent the spin and parity values of parent and daughter nuclei, respectively.
 \subsection{Nonlocality Effect}\label{B2}
\par{}
In this study, we consider the introduction of a coordinate-dependent effective mass for alpha particles. This modification of the effective mass accounts for the intrinsic nonlocal dynamical effects arising from the particle-nucleus interaction. Under these considerations, the reduced mass $\mu$ can be expressed as
\begin{equation}\label{eq14}
\mu=\frac{m^{*}M}{m^{*}+M},
\end{equation}
where $M$ is the nuclear mass of the daughter nucleus. This effective mass $m^{*}$ is expressed as the derivative of the Woods-Saxon function, multiplied by a position-dependent effective mass function \cite{teruya2016nonlocality, jaghoub2011novel, zureikat2013surface,alameer2021nucleon}. It can be written as:
\begin{equation}\label{eq15}
m^{*}=\frac{m}{1-\rho (r)},
\end{equation}
where $m$ is the free mass of $\alpha$ particle. In 2011, M. I. Jaghoub et al. \cite{jaghoub2011novel} introduced a nonlocal term into the optical model potential, which is not commonly employed for describing nucleon-nucleus elastic scattering. This nonlocal term is expressed as the derivative of the Woods-Saxon function, multiplied by a position-dependent effective mass function. In 2013, R.A. Zureikat and M.I. Jaghoub \cite{zureikat2013surface} modeled the proton-nucleus elastic scattering process by incorporating a surface term, proportional to the gradient of the target nucleus's nuclear matter density, into the traditional optical potential. These studies indicate that the phenomenological method in which a coordinate - dependent effective - mass has basically been able to accurately describe the proton-nucleus elastic scattering process. Based on these studies concerning the coordinate-dependent effective mass, it can be deduced that the $\rho(r)$ function can be defined as \cite{teruya2016nonlocality, jaghoub2011novel, zureikat2013surface,alameer2021nucleon}:
\begin{equation}\label{eq16}
\rho(r)=\rho_{S}a_{s}\frac{d}{dr}\left [ 1 + exp(\frac{r - R_{S}}{a_{S}}) \right ]^{-1}.
\end{equation}
\par{}
Where the $R_{S}$ parameter denotes the centroid location of the effective mass function $\rho$. From the research of E. L. Medeiros et al, we can know that $a_{S}$ represents the width of this function. In this work, we can obtain a set of parameters i.e., $R_{S}$, $a_{S}$, $\Delta R = 3.44$fm, which is defined as in nuclear potential $V_{N}$.
\par{}
The mass parameter $\rho_{S}$ was adjusted globally for the entire set of experimental data (see Section \ref{sec:level3}). This adjustment is necessary because it helps to better fit the experimental results and maintain the consistency of the theoretical model. By adjusting $\rho_{S}$, we can ensure that the relationship between the defined parameters like $R_{S}$, $a_{S}$, and the experimental data remains valid. In the research of E. L. Medeiros et al, we can know this adjustment occurs consistently by exchanging the free mass $m$  for its effective counterpart $m^{*}$ in the tunneling calculations.
\par
\begin{figure*}[b]
    \centering
    \includegraphics[width=\textwidth]{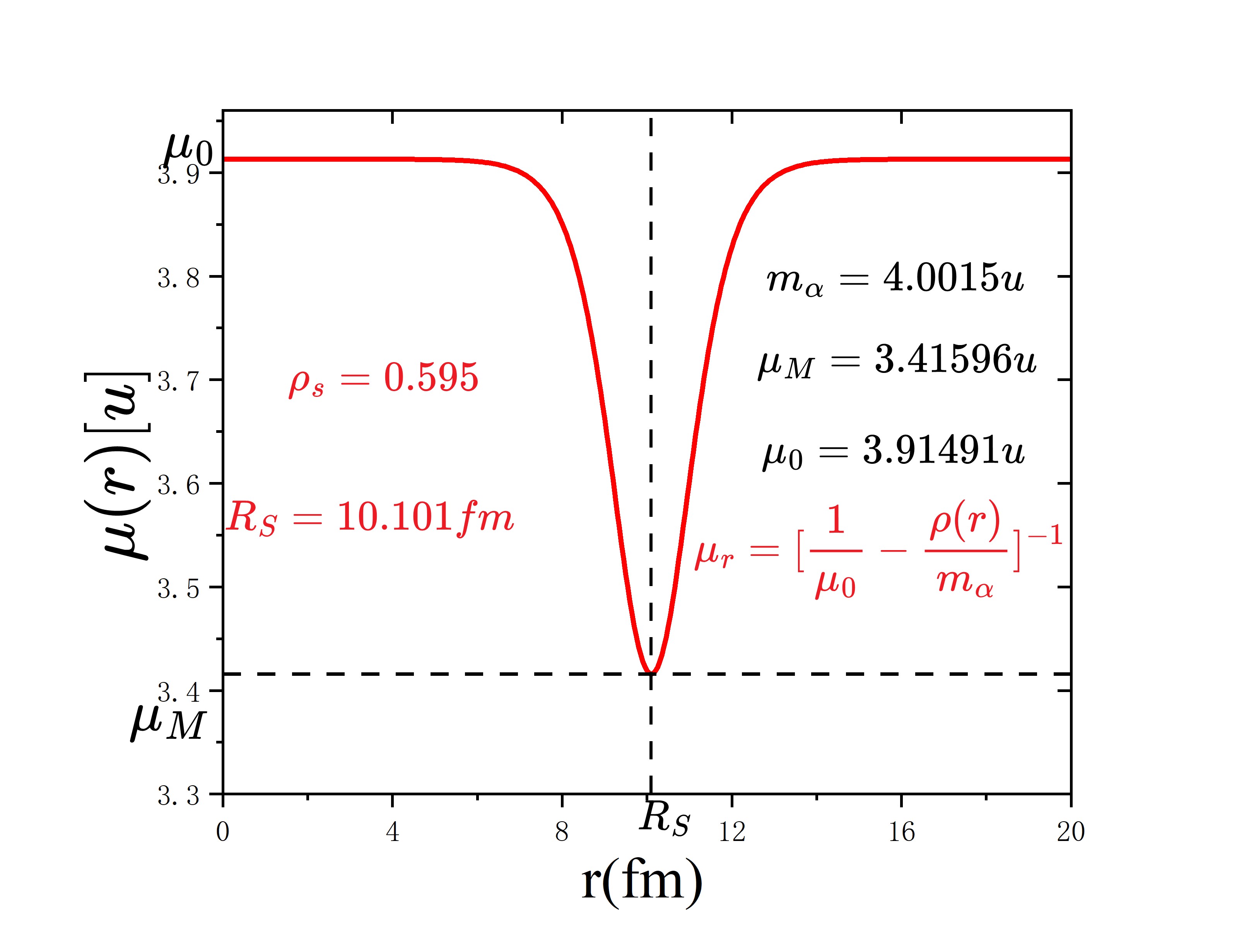}
    \caption{  The contribution of the nonlocal effect on tunneling calculations. Selected example for $\alpha$-decay from $_{78}^{184m} Pt$ : effective reduced mass $\mu$ considering nonlocality effect with $\rho_{s} = 0.595$.}
    \label{imag2}
\end{figure*}

{}
As illustrated in Fig.\ref{imag2} - \ref{imag3}, we can see the variations in the functions for the reduced effective mass $\mu$ (Fig. \ref{imag2}) and the function $f(r) = \sqrt{\mu(V - Q)}$ within the integrand in the barrier penetrability P (Eq. \ref{eq4}) during the decay process of $\alpha$-decay of  $_{78}^{184m}$ Pt. This example is presented to illustrate how  the adjusted mass parameter $\rho_{S}$ influences the calculation of barrier penetrability in a  nuclear decay scenario, further validating the theoretical framework we have established.

\begin{figure}[ht]
    \centering
    \includegraphics[width=0.5\textwidth]{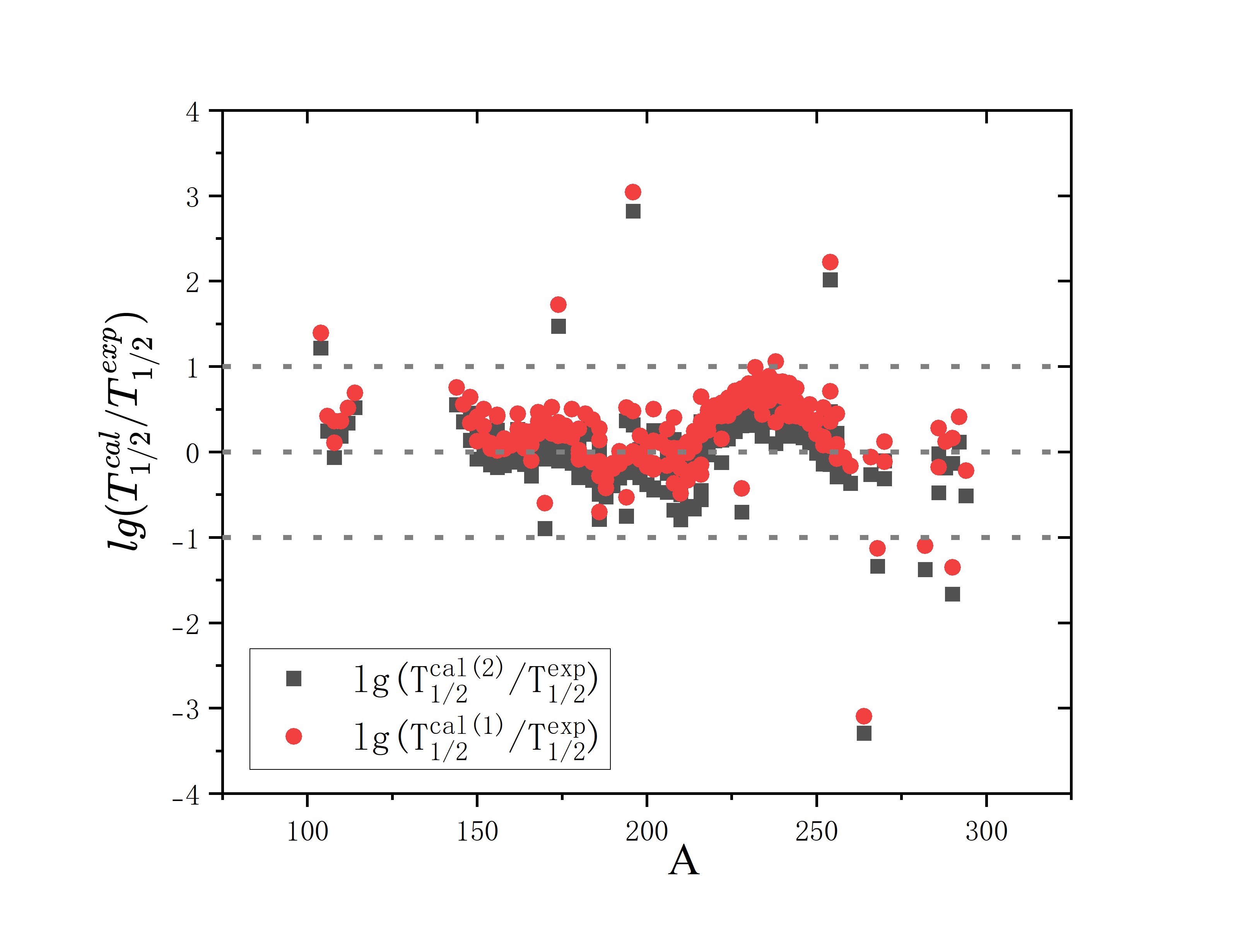}
    \caption{  The difference in logarithmic form of $\alpha$ decay half-lives between calculated data and experimental. The abscissa is the mass number A and ordiante is the value of  $log_{10}(T_{1/2}^{cal}/T_{1/2}^{exp})$. The red dot and black square represent the theoretical value calculated by using the TPA without considering nonlocality effect and after considering nonlocality effect, respectively.}
    \label{imag1}
\end{figure}

\begin{table*}[ht]
    \renewcommand{\arraystretch}{1}
    \setlength{\tabcolsep}{0.4cm}
    \centering
    \caption{The $Q_{\alpha}$ and prediction half-lives of $\alpha$ decay in even-even nuclei with Z = 118 and Z = 120. The binding energy or mass excess required to calculate the $P_{\alpha}$ and $Q_{\alpha}$ are respectively from each mass table. $Q_{\alpha}^{WS4}$, $Q_{\alpha}^{RCHB}$, and $Q_{\alpha}^{FRDM}$ indicate that the data are from mass tables WS4, RCHB, and FRDM, and the unit is MeV. $lg_{1/2}^{WS4}$ \cite{wang2014surface}, $lg_{1/2}^{RCHB}$ \cite{xia2018limits}, and $lg_{1/2}^{FRDM}$ \cite{moller2016nuclear} are the logarithms of the corresponding calculated half-life, in s.}
    \begin{ruledtabular}
    \scalebox{1}{
    \begin{tabular}{ccccccc}
        Nucleus& $Q_{\alpha}^{RCHB}$ & $Q_{\alpha}^{WS4}$&$Q_{\alpha}^{FRDM}$&$lg_{1/2}^{RCHB}$&$lg_{1/2}^{WS4}$& $lg_{1/2}^{FRDM}$ \\
        \colrule
        $^{292}118$ &10.962 & 12.237 &12.382 &-1.24&-4.48&-4.67\\
        $^{294}118$ & 10.912 & 12.195 & 12.362&-1.15&-4.40&-4.49\\
       $^{296}118$ & 10.772 & 11.748 & 12.272&-0.83&-3.41&-4.46\\
       $^{298}118$ &10.612 & 12.179 & 12.482 &-0.42&-4.44&-4.93\\
       $^{300}118$ &10.472 & 11.952& 12.502&-0.06&-3.97&-5.02\\
       $^{302}118$ &10.612 & 12.039& 12.612 &-0.33&-4.21&-5.28\\
       $^{304}118$ &12.652 & 13.118 & 13.382 &-5.45&-6.47&-6.95\\
        $^{296}120$ &11.872 & 13.339 &13.582&-2.85&-6.18&-6.48\\
       $^{298}120$ & 11.762 & 12.913  & 13.232&-2.64&-5.37&-5.92\\
       $^{300}120$ & 11.622 & 13.315 & 13.682&-2.34&-6.21&-6.79\\
        $^{302}120$ &11.512 & 12.886& 13.552& -2.10&-5.41&-6.57\\
        $^{304}120$ &11.722 & 12.759 & 13.542 &-2.48&-5.19&-6.57\\
        $^{306}120$ &13.582 & 13.776 & 14.262& -6.71&-7.17&-8.00\\
       $^{308}120$ &13.072& 12.962 & 12.962&-5.56 &-5.72&-5.81\\
    \end{tabular}
    }
    \label{tab1}
    \end{ruledtabular}
\end{table*}

\section{\label{sec:level3}RESULTS AND DISCUSSION}
At first, the experimental $\alpha$ decay half-lives $T_{1/2}^{exp}(s)$, $\alpha$ decay energies $Q_{\alpha}$ and binding energy $B(A,Z)$ are taken from the NUBASE2020 and AME2020. Recent research by E. L. Medeiros has demonstrated that by incorporating the coordinate-dependent effective mass, the theoretical outcomes of calculating the $\alpha$ decay half-lives of even-even nuclei using the semi-classical WKB method have been significantly enhanced. To further explore the application of this phenomenological approach involving the effective mass in the theoretical model of alpha decay half-lives, we employed the Two-Potential Approach (TPA) with the improvement of introducing the coordinate-dependent effective mass to calculate the $\alpha$ decay half-lives of 196 even-even nuclei. For comparative purposes, we denote the TPA without considering the effective mass as $lgT_{1/2}^{cal(1)}$ and the TPA incorporating the effective mass as $lgT_{1/2}^{cal(2)}$.
\par{}
In the TPA with the effective mass, the selection of the free mass parameter $\rho_{S}$ plays a crucial role. To minimize the discrepancy between the improved TPA (i.e., the calculation results considering the effective mass) and the experimental data, the adjustment $\rho_{S}$ criterion for the parameters is set to reduce the differences between the final calculation results of all even-even nuclei and the experimental results. In essence, this aims to decrease the standard deviation between the calculated values and the experimental data. Consequently, the standard deviation $\sigma _{\rho_{S}}$ at this stage can be expressed as:
\begin{equation}\label{eq17}
\sigma _{\rho_{S}}=\sqrt{\frac{1}{n}\sum_{i=1}^{n}(\Delta_{i})^{(2)}};\ \Delta_{i}= log_{10}(T_{i}^{cal})-log_{10}(T_{i}^{exp}),
\end{equation}
where $T_{i}^{cal}$ and $T_{i}^{exp}$ are the calculated and experimental half-lives of the i-th decaying nucleus, respectively, and $\Delta_{i}$ denotes the logarithmic deviation between calculated and experimental data. Compared with the experimental results, when the value of the adjustment $\rho_{S}$ is set to 0, its standard deviation $\sigma_{\rho_{S} = 0}$ = 0.573, but when the value of the adjustment $\rho_{S}$ is set to 0.595, its standard deviation  $\sigma_{\rho_{S} = 0.595}$ = 0.522. These results indicate that when the coordinate-dependent effective mass is considered, that is, when the adjustment $\rho_{S}$ is set to 0.595, the calculation results of TPA have been significantly improved. It validates the importance of incorporating the nonlocality effect in our study and provides a more reliable theoretical framework for understanding the $\alpha$ decay process of even - even nuclei. Further emphasizing that now the standard deviation has increased by approximately $9\%$.
\par
\begin{figure*}[b]
    \centering
    \includegraphics[width=0.7\textwidth]{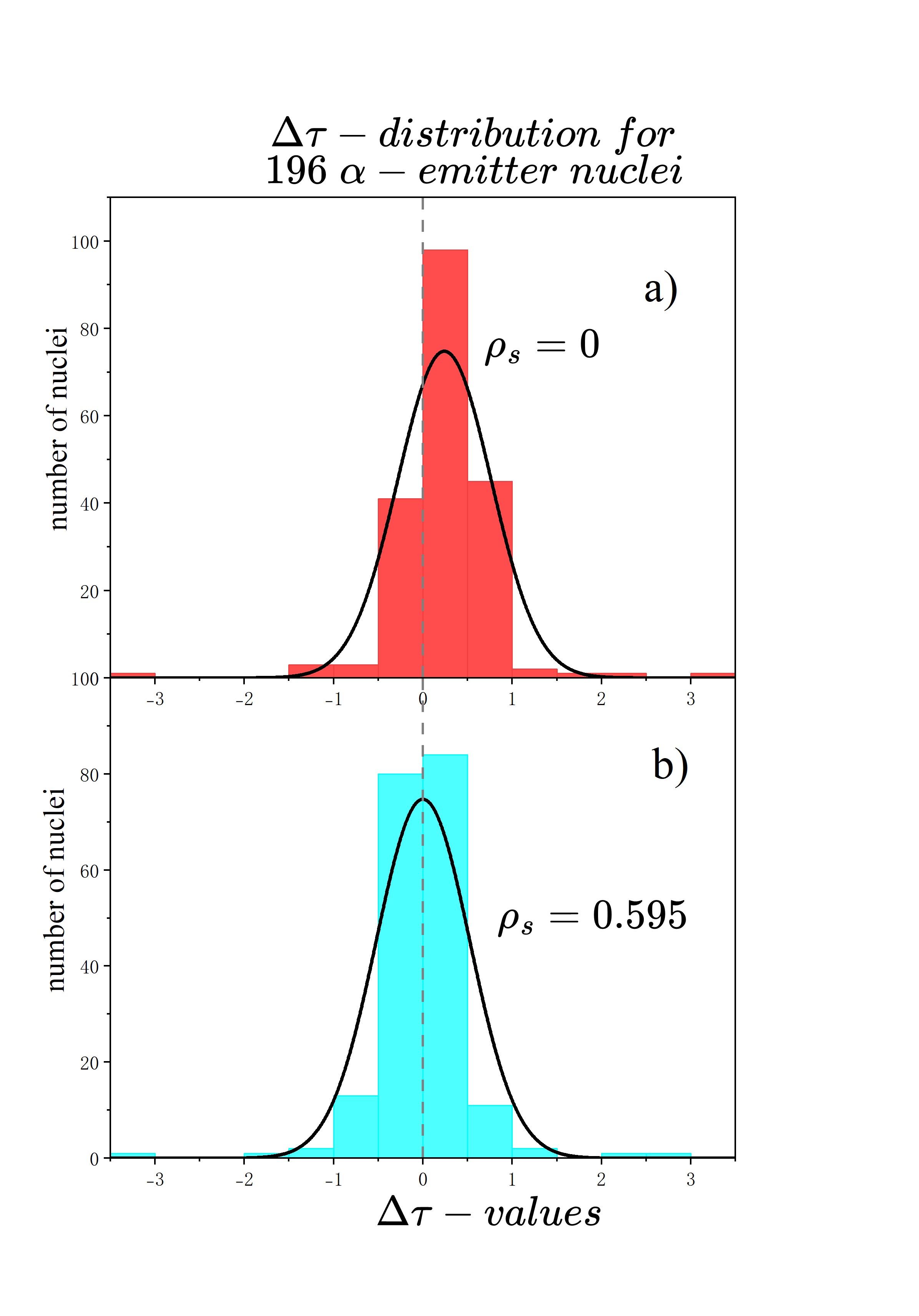}
    \caption{  $\Delta \tau$-distributions for 196 $\alpha$-emitters. (a) Result for the calculations with $\rho_{s} = 0$ (without nonlocality effect). In this case, most of the calculated half-lives are bigger than the experimental values, which can be observed with the centroid being shifted to the positive values $\Delta \tau$ = 47.232. (b) The centroid is exactly on $\Delta \tau$ = 0 when the nonlocality effect is considered in the calculation with $\rho_{s}$ = 0.595.}
    \label{imag4}
\end{figure*}

{}
The Fig. \ref{imag4} clearly illustrate the impact of the nonlocality effect on the calculations.  Fig. \ref{imag4}a reveals that, in the absence of considering the nonlocal effect, the centroid of the distribution shifts in the positive direction. This observation indicates that the results obtained by the TPA without considering the effective mass predominantly exceed the experimental values under such circumstances. To align the centroid with zero, thereby minimizing the standard deviation, we set the adjustment $\rho_{S}$ to 0.59. The centroid distribution depicted in Fig. \ref{imag4}b further validates the correctness of our approach.
\par
\begin{figure*}[b]
    \centering
    \includegraphics[width=\textwidth]{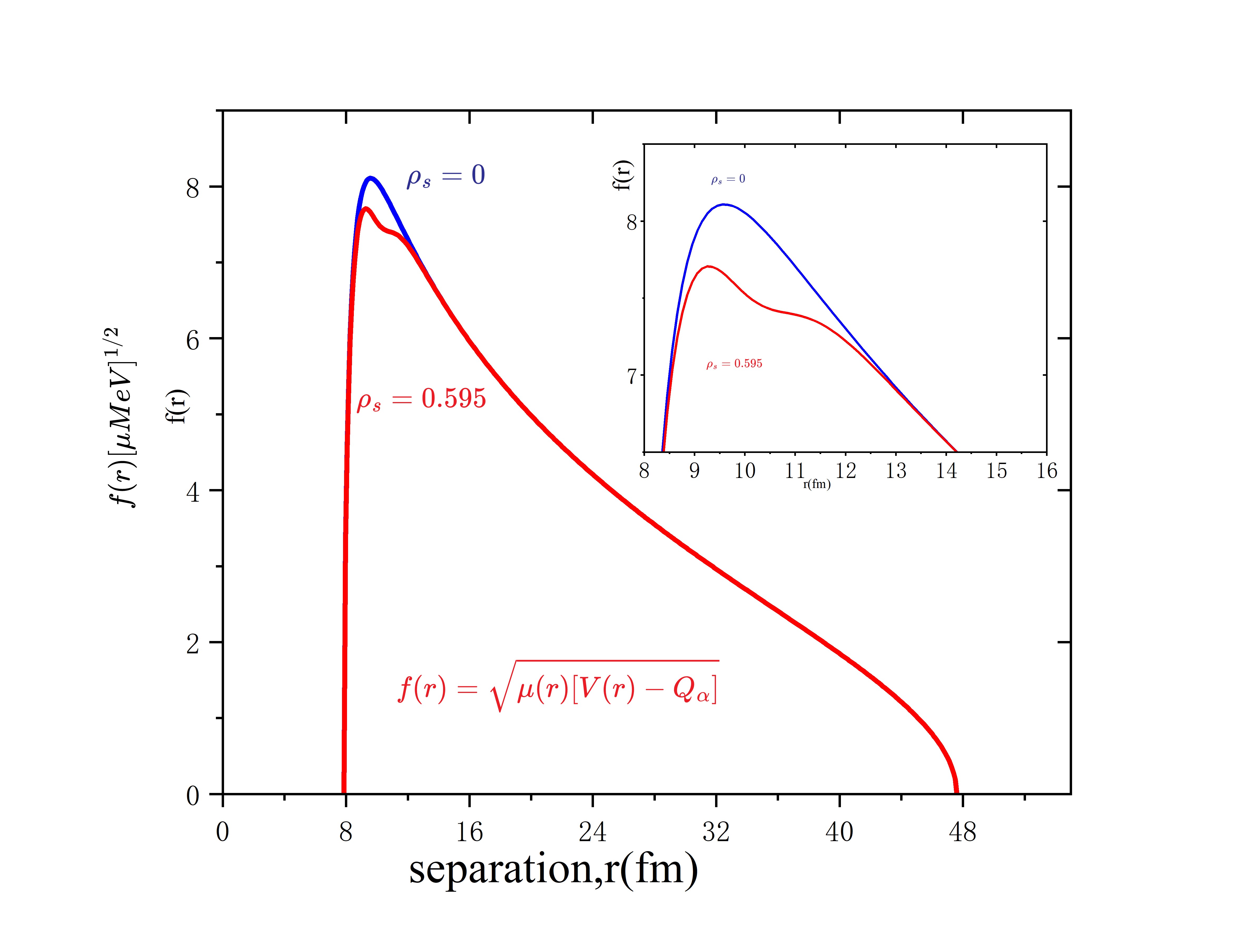}
    \caption{   The contribution of the nonlocal effect on tunneling calculations. Selected example for $\alpha$-decay from $_{78}^{184m} Pt$ : comparison between the functions $f(r)$ in the integrand of the barrier penetrability: considering the reduced masses $\mu$ (blue line $\rho_{S} = 0$ and $\mu_{0}$ (red line $\rho_{S} = 0.595$)).}
    \label{imag3}
\end{figure*}

{}
Finally, we make a summary of the contribution produced by the dynamic effect of the non-locality of potential. It can be seen from Fig. \ref{imag2} - \ref{imag3} that when the non-locality of the potential is considered, its effective mass will decrease. Specifically, at $r = R_{S}$, its effective mass decreases by approximately $13\%$. These results are all in line with those of E. L. Medeiros et al.
\par{}
As an application, we extend our improved TPA model to predict the $\alpha$ decay half-lives of 14 even-even nuclei with Z = 118 and Z = 120 in the following. Moreover, given the high sensitivity of the half-life to $Q_{\alpha}$, it is essential to choose an appropriate mass model for calculating $Q_{\alpha}$. In this study, for the purpose of comparison, we employed three mass tables i.e. WS4 \cite{wang2014surface}, relativistic continuum Hartree-Bogoliubov (RCHB) \cite{xia2018limits} and FRDM \cite{moller2016nuclear}, to obtain $Q_{\alpha}$. The prediction results are presented in Table 5. In this table, the first column is $\alpha$ decay parent nucleus, and the second to fourth columns are the different $Q_{\alpha}$ from the three mass tables, respectively. The last three columns are the logarithmic form of $\alpha$ decay half-lives corresponding to the three mass tables. Additionally, to facilitate a more comprehensive understanding of how these three mass models affect the half-life calculations, we plot the predicted half-lives of isotope chains for Z = 118 and Z = 120 in figure \ref{imag5}. From this figure, we can clearly see that N = 184 has an obvious shell effect. However, while the shell effect of N = 184 is clearly visible in all cases,the shell effect of N = 178 is only observed in the WS4 \cite{wang2014surface} and the FRDM \cite{moller2016nuclear} models, and not in the RCHB  \cite{xia2018limits} model. It is not difficult to draw a conclusion: N = 184 is a magic number, but the shell effect of N = 178 depends on the models used.
\section{\label{sec:level4}SUMMARY AND CONCLUSION}
\begin{figure*}[b]
    \centering
    \includegraphics[width=0.7\textwidth]{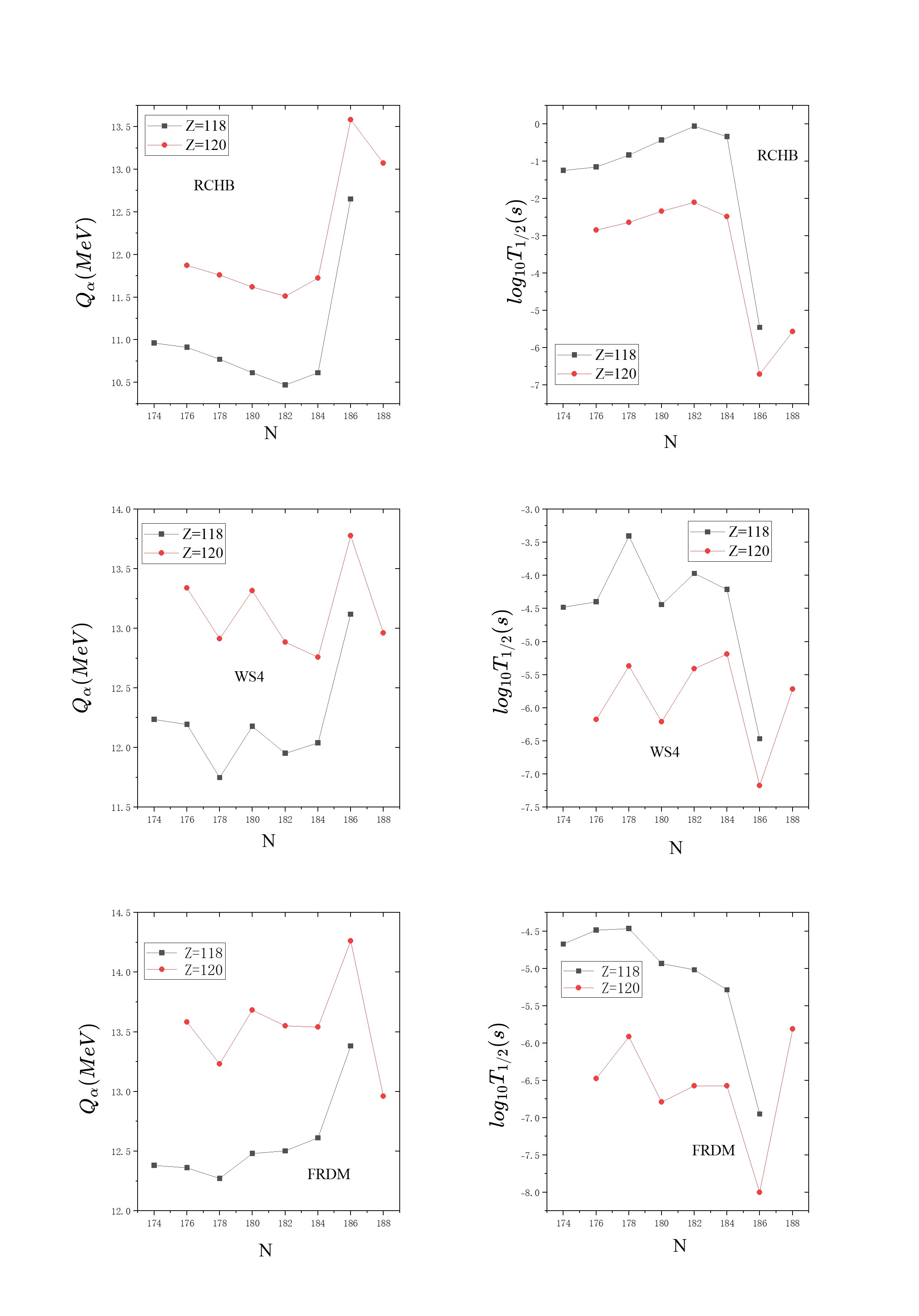}
    \caption{ The value of $Q_{\alpha}$ and predicted $\alpha$ decay half-lives for even-even nuclei with Z = 118 and 120 isotopes. The black square and red dot indicate Z = 118 and Z = 120, respectively. The abscissa is neutron number $N$, the ordinate in the left column is $Q_{\alpha}$ in MeV, and the ordinate in the right column is logarithm $log_{10}T_{1/2}$ of calculated half-life in s. The mass tables used from top to bottom in this figure are RCHB, WS4, FRDM.}
    \label{imag5}
\end{figure*}

In summary, we conducted a systematic calculation of the Two-Potential Approach (TPA) of A by introducing the coordinate-dependent effective mass for the alpha particle, which reflects the nonlocality effective of the particle-nucleus interaction. Our findings indicate that this improved TPA demonstrates a closer agreement with the experimental results compared to the TPA that the nonlocality effective of the of the particle-nucleus interaction. As an application, we utilized the decay energies calculated by the three mass tables of WS4, RCHB, FRDM as input values and employed the improved TPA to predict the $\alpha$ decay half-lives of 14 even-even nuclei with Z = 118 and Z = 120 isotopes. From Fig. \ref{imag5}, we can conclude that N = 184 is a magic number and N = 178 is dependent on models. This research can offer valuable insights for the study of new elements and isotopes in future experimental investigations.


\bibliography{study}
\end{document}